\documentclass[conference]{IEEEtran}
\IEEEoverridecommandlockouts
\usepackage{cite}
\usepackage{amsmath,amssymb,amsfonts}
\usepackage{algorithmic}
\usepackage{graphicx}
\usepackage{textcomp}
\usepackage{xcolor}
\usepackage{xcolor}
\usepackage{url}
\usepackage{multirow}
\usepackage{tabularx}

\def\BibTeX{{\rm B\kern-.05em{\sc i\kern-.025em b}\kern-.08em
    T\kern-.1667em\lower.7ex\hbox{E}\kern-.125emX}}
\begin{document}

\title{On Medical Device\\ Cybersecurity Compliance in EU
}

\author{\IEEEauthorblockN{Tuomas Granlund}
\IEEEauthorblockA{\textit{Solita Oy \& University of Tampere} \\
Tampere, Finland \\
tuomas.granlund@solita.fi}
\and
\IEEEauthorblockN{Juha Vedenp\"{a}\"{a}}
\IEEEauthorblockA{
\textit{Solita Oy} \\
Tampere, Finland \\
juha.vedenpaa@solita.fi}
\and
\IEEEauthorblockN{Vlad Stirbu}
\IEEEauthorblockA{
\textit{CompliancePal}\\
Tampere, Finland \\
vlad.stirbu@compliancepal.eu}
\and
\IEEEauthorblockN{Tommi Mikkonen}
\IEEEauthorblockA{
\textit{University of Helsinki}\\
Helsinki, Finland \\
tommi.mikkonen@helsinki.fi}
}

\maketitle

\begin{abstract}
The medical device products at the European Union market must be safe and effective. To ensure this, medical device manufacturers must comply to the new regulatory requirements brought by the Medical Device Regulation (MDR) and the In Vitro Diagnostic Medical Device Regulation (IVDR). In general, the new regulations increase regulatory requirements and oversight, especially for medical software, and this is also true for requirements related to cybersecurity, which are now explicitly addressed in the legislation. The significant legislation changes currently underway, combined with increased cybersecurity requirements, create unique challenges for manufacturers to comply with the regulatory framework. In this paper, we review the new cybersecurity requirements in the light of currently available guidance documents, and pinpoint four core concepts around which cybersecurity compliance can be built. We argue that these core concepts form a foundations for cybersecurity compliance in the European Union regulatory framework. 
\end{abstract}

\begin{IEEEkeywords}
Medical device, regulatory requirements, cybersecurity, regulatory compliance
\end{IEEEkeywords}

\section{Introduction}

Regulatory compliance is one of the most important quality characteristics of a medical device -- placing such device on the European Union (EU) market requires that the device complies with the EU regulatory framework. Also, the processes by which the device is being manufactured and maintained must be compliant with the regulations. In addition to medical device regulations, the manufacturer may need to comply with some other specific European legislation, depending on the product type. 

At present, the EU medical device legislation is in transition, where  three former directives will be replaced by two new regulations. A shared understanding of the contents amongst the practitioners is continuously being refined. This takes place in many ways, including new harmonized standards and guidance documents. In general, the transition is resulting in an increase in requirements, and a vital example of this is the more stringent requirements for cybersecurity. Furthermore, cybersecurity is a topical issue due to the acceleration of the digital transformation in health care and the increasing complexity of the new devices.  

In this paper, we address the new cybersecurity requirements from a high viewpoint and propose an approach that can be used to build an understanding of the most fundamental aspects of cybersecurity compliance. It is crucial to fully understand the new requirements to implement them efficiently. This requires in-depth knowledge of practical methods and tools. 

The rest of this paper is structured as follows. In Section \ref{background}, we provide the legal background for the paper. In Section \ref{common_challenges}, we address some apparent common challenges related to cybersecurity compliance concerning the new regulatory requirements. In Section \ref{foundations}, which forms the paper's core, we present our proposal for \textit{Foundations of Medical Device Cybersecurity Compliance}. In Section \ref{callforaction}, we present a call for action for both regulatory authorities and medical device manufacturers. Finally, the concluding remarks are provided in Section \ref{conclusions}. 

\section{Background}
\label{background}
Currently, the EU medical device legislation is undergoing a significant change. Three former directives will be replaced by two new regulations: the Medical Device Regulation (MDR) 2017/745\cite{mdr} and the In Vitro Diagnostic Medical Device Regulation (IVDR) 2017/746 \cite{ivdr}. To address certain identified shortcomings of the former legislation, the new regulations emphasize the importance of clinical evaluation, risk management, post-market surveillance activities, and risk-benefit analysis throughout the product lifecycle \cite{pitkanen2020}. In addition, particular new concepts are introduced, including now explicitly addressed requirements related to cybersecurity. 

The increased cybersecurity requirements in the MDR and the IVDR affect mainly physical medical device products containing software and software that are the device itself. Also, specific requirements connected to the device's operational environment are presented, for example, requirements related to IT network characteristics. 

In addition to MDR and IVDR, both of which are legally binding legislative acts and must be applied in their entirety, the EU Commission has mechanisms to provide more detailed technical specifications, in the form of harmonized standards and a range of non-binding guideline documents, for instance. The standards can be used to address General Safety and Performance Requirements (Annex I of MDR or IVDR), and guidance documents are intended to support the uniform application of the regulations within the EU. 

Related to cybersecurity requirements, currently, the most important source for information is MDCG guidance document \textit{MDCG 2019-16, Guidance on Cybersecurity for medical devices}\cite{mdcg_2019-16}. At present, there are no harmonized standards against MDR or IVDR, which creates uncertainty for manufacturer concerning the appropriate set of standards to be selected to provide a 'presumption of conformity' against the regulatory requirements. However, the EU commissions recent draft standardization request \cite{draft_request} could be used to get an insight into the expectations of regulatory authorities related to applicable standards. The draft request includes a forthcoming standard \textit{IEC 80001-5-1} \cite{iec80001-5-1}, which is closely related to cybersecurity and relevant for the manufacturers. In addition to the sources mentioned above, the guidance documents \textit{AAMI TIR57:2016 Principles for medical device security—Risk management}\cite{aami_tir57} and \textit{IMDRF Principles and Practices for Medical Device Cybersecurity} \cite{imdrf_2019} have been used as a reference in this paper. 

\section{Common challenges for the manufacturers}
\label{common_challenges}

Based on the industry experiences, even the former regulatory requirements create certain challenges for the manufacturers \cite{williams2015}. While safety risks emerging from the cybersecurity dimension should be appropriately managed already today, some companies seem to have shortcomings in their security culture and processes \cite{ansm2019}. Therefore, the gap between the new requirements and the existing implementation can be more significant than the actual difference between the existing and the new legislation is.

The EU regulatory framework is characterized by a certain level of complexity, which is true for cybersecurity requirements as well: the requirements in the new regulations concern many interconnected processes, and as a result, they are divided into different sections of the document. Therefore, the initial challenge for compliance is identifying and understanding all applicable provisions of the regulations. Furthermore, manufacturers have challenges addressing cybersecurity aspects within their risk management processes and implementing security actions into their design and development processes. It is evident that manufacturers need resources with in-depth knowledge about cybersecurity to deal with the new threats arising from accelerating digital transformation. 

\section{Towards Medical Device Cybersecurity Compliance}
\label{foundations}

In this section, we propose a high-level approach to address the new legislation's cybersecurity requirements. Special attention has been paid to the efficiency, making use of existing process requirements related to medical device design and development. 

\subsection{Risk relationship between safety and security}

Medical device manufacturers are required to manage the risks related to their products throughout the devices' whole life cycle \cite{mdr}, \cite{ivdr}, \cite{imdrf_2019}. In practice, medical software risk management activities are implemented according to the requirements of \textit{ISO 14971}\cite{iso14971} and \textit{IEC 62304}\cite{iec62304}. In ISO 14971, the definition of risk is tightly bound to the concept of "physical injury or damage to the health of people, or damage to property or the environment." The risk management approach derived from this definition is characterized by the fact that its perspective focuses solely on safety-related risks, i.e., risks that affect the safety of the patients or the users. However, it is evident that the new regulations require a broader view of risk to address the new cybersecurity requirements. 

\begin{figure}[t]
    \centering
    \includegraphics[width=0.43\textwidth]{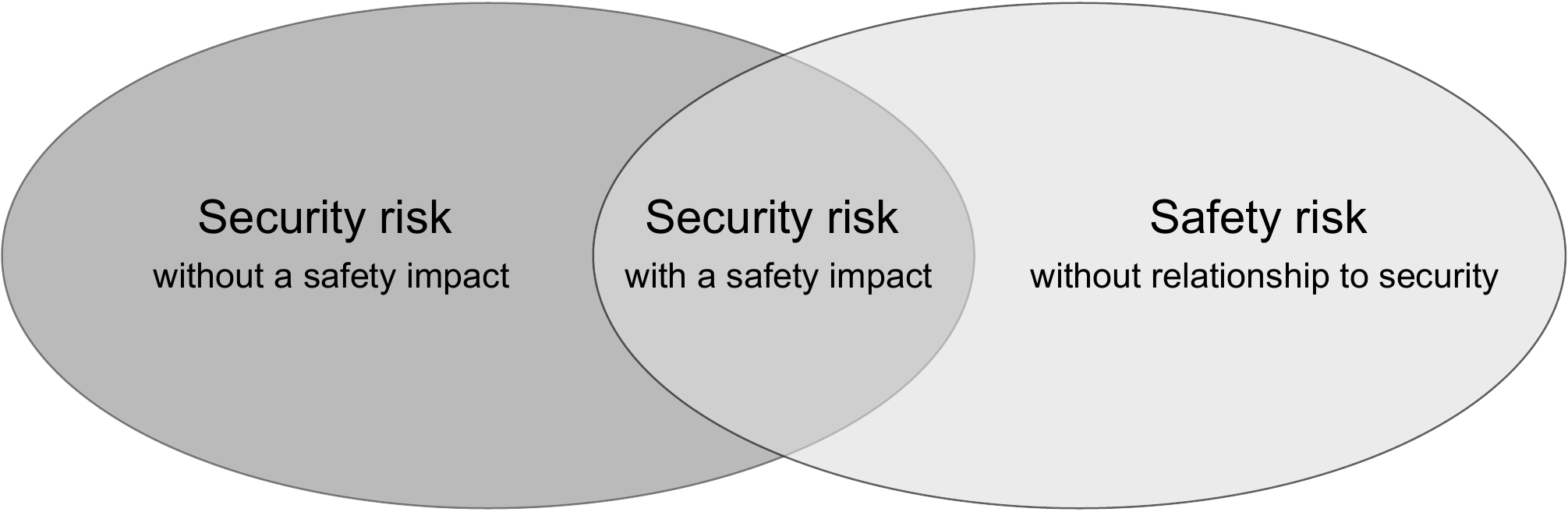}
    \caption{Risk relationship between safety and security. Adapted from MDCG 2019-16, Guidance on Cybersecurity for medical devices.}
    \label{fig:risk_relationship}
\end{figure}

The relationship between different risk classes is well documented \cite{mdcg_2019-16}, \cite{aami_tir57}, \cite{ansm2019} \cite{iec_63069} (Figure \ref{fig:risk_relationship}). In the context of cybersecurity, there are three main types of risks: 
\begin{itemize}
    \item Security risk without a safety impact,
    \item Security risk with a safety impact,
    \item Safety risk without relation to security.
\end{itemize}
MDCG 2019-16 further elaborates the concept of "the relationship between cybersecurity risk management and patient safety management" by providing a table of examples in Annex II, from which more detailed risk types can be gathered. According to MDCG 2019-16, there can also be security risk controls with safety impacts and security risks with indirect safety impacts (e.g., risks related to device availability). When considering software-only devices, the safety risks are indirect by nature (i.e., incorrect or delayed diagnosis or treatment) \cite{imdrf_2014}, whereas physical devices can directly hurt the patient. 

\subsection{Aligned safety and security risk management processes}

A safety risk management process can be considered as one of the core processes of medical device manufacturing \cite{pitkanen2020}. Therefore, it is essential that the risk management process is working efficiently. When cybersecurity aspects are also considered, an aligned and efficient risk management process can be built utilizing the above-defined safety-security risk relationship as the baseline.

\begin{figure}[h]
    \centering
    \includegraphics[width=0.43\textwidth]{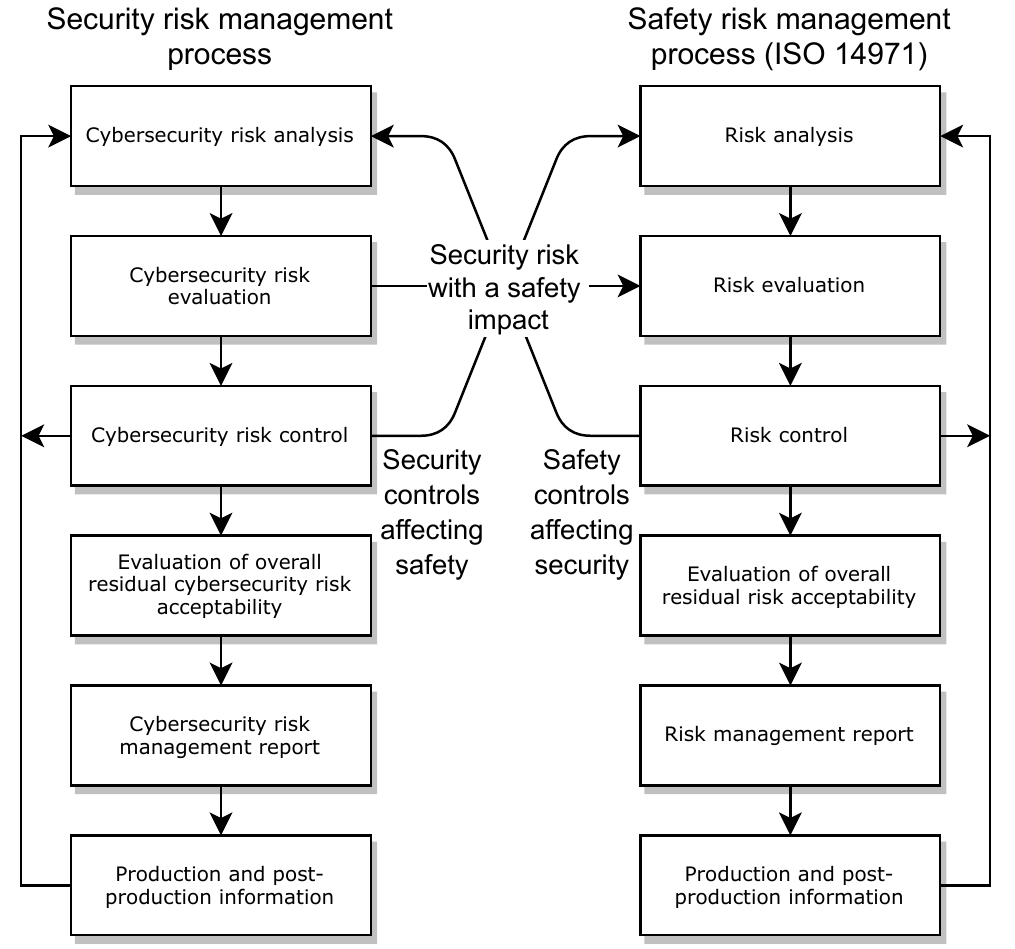}
    \caption{Aligned safety and security risk management processes. Adapted from MDCG 2019-16, Guidance on Cybersecurity for medical devices and AAMI TIR57:2016, Principles for medical device security--Risk management.}
    \label{fig:aligned_processes}
\end{figure}

A security risk management process can be presented as a companion process to the safety risk management process (Figure \ref{fig:aligned_processes}), similarly to the usability engineering process in \textit{IEC 62366-1}  \cite{aami_tir57}, \cite{iec62366-1}. Although not explicitly required \cite{mdcg_2019-16}, a separate security risk management process is recommended due to a broader perception of risk and to ensure complete and consistent security management \cite{aami_tir57} -- the two processes differ in accordance to the nature of the risks that they control \cite{ansm2019}. The security risk management process is used to manage those security risks that do not contain direct or in-direct safety impacts as they need to be managed to fulfill the cybersecurity requirements in MDR and IVDR. However, if a security risk has a safety impact, it will be propagated as an input item to safety risk evaluation as required by ISO 14971. Furthermore, as new risk control measures may introduce new risks, the impact of control must be analyzed from both safety and security perspectives.

MDCG 2019-16 has recognized potential challenges related to security risk controls -- both too weak and too restrictive security controls may lead to undesired safety impact \cite{mdcg_2019-16}. As a result, it is essential to ensure that the separate processes are well-aligned to enable the iteration of the device design between the processes through their logical connection points discussed above. Transparency is needed to set the balance between safety and security correctly and to address potential trade-offs explicitly. 

\subsection{Secure design and development lifecycle}

As discussed above, security risk management is an integral part of the secure development lifecycle. One of the first steps of secure development is to perform threat modeling to identify potential security threats towards the device in its operating environment \cite{mdcg_2019-16}, \cite{iec80001-5-1}. In-depth security expertise is needed to capture all relevant security aspects fully - similarly as clinical expertise is needed in safety risk identification. Software security requirements are derived from the threat model to be implemented as appropriate risk controls. 

Similarly to risk management being implemented in accordance to ISO 14971, medical software manufacturers generally have their product development processes implemented in compliance to the requirements of ISO 13485 and IEC 62304. From this point of view, it is recommended to extend the existing development processes with cybersecurity activities. This approach is supported by the forthcoming standard IEC 80001-5-1 \cite{iec80001-5-1}. Furthermore, the next version of IEC 62304 will likely also include specific cybersecurity provisions \cite{iec62304_ed2},  \cite{varri2019}. 

MDCG 2019-16 approaches secure design and manufacturing process through "defense in depth" strategy. The strategy consists of the eight most necessary practices, which are no different from other cyber-physical systems \cite{mdcg_2019-16}. The practices are presented in Table \ref{table:1}. In addition, Table \ref{table:1} shows how the practices can be mapped to IEC 80001-5-1 processes and requirements. As can be seen, the approaches taken by the documents are well-aligned, and IEC 80001-5-1 provides a more pragmatic view for implementing required security concepts. Therefore, IEC 80001-5-1 can be considered as a welcomed addition to the set of applicable medical device standards in the future.

\begin{table}[t]
\caption{Correspondence between MDCG 2019-16 Defence-in-depth practices and IEC 80001-5-1 processes.}
\label{table:1}
\begin{tabularx}{0.48\textwidth} { 
  | >{\raggedright\arraybackslash}X 
  | >{\raggedright\arraybackslash}X | }
 \hline
MDCG 2019-16 & IEC 80001-5-1  \\ 
 \hline
\multirow{2}{12em}{1. Security management} 
& 4.1 Quality Management \\ 
& 5.1 Software Development Planning \\ 
\hline
\multirow{2}{12em}{2. Specification of security requirements}
& 5.2 Health Software Requirement Analysis \\
\hline
\multirow{2}{12em}{3. Secure by design}
& 5.3 Software Architectural Design \\
& 5.4 Software Detailed Design \\
\hline
\multirow{2}{12em}{4. Secure implementation}
& 5.1 Software Development Planning \\
& 5.5 Software Unit Implementation and Verification \\
& 5.8.3 File Integrity \\
& 5.8.4 Controls for private keys \\
\hline
\multirow{2}{12em}{5. Security verification and validation testing}
& 5.5 Software Unit Implementation and Verification \\
& 5.6 Software Integration Testing \\
& 5.7 Software System Testing \\
& 5.8 Software Release \\
\hline
\multirow{2}{12em}{6. Management of security-related issues}
& 4.1 Quality Management \\ 
& 5.1 Software Development Planning \\ 
& 6 Software Maintenance Process \\ 
& 7 Security Risk Management \\
& 8 Software Configuration Management Process \\
& 9 Software Problem Resolution Process \\
& 10 Quality Management System \\
\hline
\multirow{2}{12em}{7. Security update management}
& 6.1 Security Update Management \\
& 6.2 Implementation Policy \\
& 6.3 Post-Market activities for Health Software \\
\hline
\multirow{2}{12em}{8. Security guidelines}
& 5.8.2 Release documentation \\
& 6.3.1 Security update documentation \\
& 10.6 Accompanying Documents Review \\
\hline
\end{tabularx}
\end{table}

\subsection{Post-market cybersecurity activities}

The new regulations require a more proactive approach to post-market surveillance from manufacturers. In general, the number of safety-related hazards of a medical device will stay relatively stable over time  \cite{mdcg_2019-16}. However, the same does not hold when considering cybersecurity threats, and, as a consequence, post-market cybersecurity activities need special consideration. 

Manufacturers are required to have post-market surveillance (PMS) system in place, which includes, depending on the class of the device, PMS reports, or Periodic Safety Update Reports (PSUR) \cite{mdr}, \cite{ivdr}. These reports should summarize the results and conclusions of the analysis of all the data from the market \cite{mdcg_2019-16}, including cybersecurity observations. Self-evidently, relevant changes in the security environment  should be addressed as an input to the risk management process in a more timely manner. In addition, the vigilance process managing serious incidents and field safety corrective actions should be fine-tuned to consider also security aspects.  

The key to efficient cybersecurity post-market surveillance process is identifying new security vulnerabilities proactively and resolving them timely yet without compromising safety or compliance. In addition to active vulnerability monitoring, manufacturers need a formal vulnerability disclosure process to communicate vulnerabilities with stakeholders and update technical documentation accordingly. In addition, the manufacturer must have a clear plan for recovery following cybersecurity incidents \cite{imdrf_2019}, also considering safe and secure decommissioning of the device.

\section{Call for action}
\label{callforaction}

We believe that there is an urgent need for a call for action for regulatory authorities and medical device manufacturers. In the absence of harmonized standards against the MDR or IVDR, the crucial concept of 'presumption of conformity' is malfunctioning \cite{pitkanen2020}, creating unnecessary problems for the manufacturers. Combined with the complexity level of the regulatory framework and somewhat latent MDCG cybersecurity guidance, there is an evident shortcoming in the shared understanding of regulatory authorities' expectations related to cybersecurity requirements. We argue that IEC 80001-5-1 is an appropriate standard to address the cybersecurity regulatory requirements, and for this reason, its use should be adopted by the regulatory authorities and the manufacturers without undue delay. 

In parallel, the manufacturers need to improve their underdeveloped culture of cybersecurity and set up their design and development processes according to the new requirements. Our proposed approach for the foundations of cybersecurity compliance serves as a good baseline for the implementation.
\section{Conclusions}
\label{conclusions}

Compliance with the regulatory requirements is one of the most important yet challenging aspects of medical device manufacturing. The new EU medical device legislation contains increased cybersecurity requirements, and the interconnection between the different documents, document sections, and requirements can be a challenge for the manufacturers. With regard to the current legislation transition and cybersecurity requirements, we identified the most essential sources for regulatory requirements and guidelines currently available. 

With this paper, we proposed an approach that can be used to understand cybersecurity compliance and explain the key concepts. First, the foundations of medical device cybersecurity compliance are built on the risk relationship between safety and security. Second, an effective risk management approach can be derived from separate yet well-aligned safety and security risk management processes. Third, cybersecurity is implemented in practice within a well-controlled secure design and development process.  Finally, medical devices' cybersecurity aspects must be considered throughout the whole lifecycle of the device. 

Our research emphasized cross-analysis of the secure design and manufacture requirements of MDCG 2019-16 and IEC 80001-5-1. The practical outcome of this analysis is a table that shows the correspondence between the process requirements. Our results imply that forthcoming standard IEC 80001-5-1 provides an appropriate information source for manufacturers on how to implement the cybersecurity regulatory requirements in practice.

\textbf{Acknowledgment}.
The authors would like to thank Business Finland and the members of AHMED (Agile and Holistic MEdical software Development) consortium for their contribution to preparing this paper.

\end{document}